\documentclass[twocolumn,dvipdfmx,shortnote]{jpsj3}
\usepackage[dvipdfmx]{graphicx}
\usepackage{amsmath,amssymb}
\usepackage{bm}
\def\vector#1{\mib #1}

\newcommand{\entropy}{\mathcal{S}}
\newcommand{\heat}{\mathcal{C}}

\newcommand{\ket}[1]{\left\vert {#1} \right\rangle}
\def\Nd{N_{\rm d}}
\def\Ndw{{N_{\downarrow}}}

\def\theenumi{{\roman{enumi}}}
\newcommand{\aver}[1]{\left\langle {#1} \right\rangle}
\title
{Exact Thermodynamic Properties of $(1,1/2)$ Mixed Diamond Chains with Strong Single-Site Anisotropy}

\author
{Ryuta Iwazaki and Kazuo Hida\thanks{E-mail address: hida@mail.saitama-u.ac.jp}
}

\inst
{Division of Material Science, Graduate School of Science and Engineering, \\ Saitama University, Saitama, Saitama, 338-8570}

\recdate
{
\today
}

\abst{
The ground states and {finite-temperature} properties of mixed diamond chains with spins 1 and 1/2 are investigated in the limit of strong easy-axis anisotropy on the spin-1 sites.
Magnetization curves, entropy, specific heat and magnetic susceptibility are exactly calculated using the method of {\v{C}anov\`a} {\it et al.} [J. Phys.: Condens. Matter {\bf 18} 4967 (2006)].
}

\begin{document}
\sloppy
\maketitle

We consider the mixed diamond chains with single-site anisotropy {$D$ and magnetic field $H$} described by the following Hamiltonian:
\begin{align}
{\cal H} =& \sum_{l=1}^{L} \left[
(\vector{S}_{l}+\vector{S}_{l+1})\cdot(\vector{\tau}^{(1)}_{l}+\vector{\tau}^{(2)}_{l})+ \lambda\vector{\tau}^{(1)}_{l}\vector{\tau}^{(2)}_{l}\right]
\nonumber\\
&+D\sum_{l=1}^{L}S^{z2}_l-H\sum_{l=1}^{L}\left(\tau^{(1)z}_l+\tau^{(2)z}_l+S^z_{l}\right),
\label{ham0}
\end{align}
where $\vector{S}_{l}$ is a spin-1 operator, and
$\vector{\tau}^{(\alpha)}_{l} (\alpha=1,2)$ are spin-1/2 operators in the $l$th unit cell.
The total number of unit cells is denoted by $L$.
The ground state and finite temperature properties of this model have been investigated by one of the authors and coworkers\cite{hida0,hida1}.
Defining the composite spin operators  $\vector{T}_{l} \equiv \vector{\tau}^{(1)}_{l}+\vector{\tau}^{(2)}_{l}$, it is  evident that ${\forall l}\ [\vector{T}_l^2, {\mathcal H}] = 0$.
Thus, we have $L$ conserved quantities $\vector{T}_l^2(\equiv T_l(T_l+1);\ T_l=0 $ or 1).
The total Hilbert space of the Hamiltonian (\ref{ham0}) consists of separate subspaces, each of which is specified by a definite set of $\{T_l\}$.

A spin pair with $T_l=0$ is a singlet dimer that cuts off the correlation between $\vector{S}_{l}$ and  $\vector{S}_{l+1}$. 
 The segment including $n$ successive $\vector{T}_{l}$'s  with $T_l=1$ and $n+1$ $\vector{S}_{l}$'s coupled with them
 is called a cluster-$n$.
A  cluster-$n$ is equivalent to a spin-1 antiferromagnetic Heisenberg chain of length $2n+1$  with alternating single-site anisotropy.
 A dimer-cluster-$n$ (DC$n$) phase  consists of an alternating array of cluster-$n$'s and dimers.

Since all eigenstates are constructed as direct products of the eigenstates of cluster-$n$'s and dimers, the full thermodynamics of diamond chains can be formally formulated.
However, the calculations of thermodynamic quantities include the contribution from cluster-$n$'s of arbitrary size.
This is intractable in practice.
In the region where the ground state is a DC$n$ phase with finite $n$, the contributions from large cluster-$n$'s are small.
Hence, the thermodynamic properties can be estimated with enough accuracy\cite{hida0,hida1}.
On the other hand, in the region where the ground state is a single infinite-size cluster (DC$\infty$ phase), this approximation is unreliable.

In the present paper, we consider the easy-axis limit ($D \rightarrow -\infty$) of the Hamiltonian (\ref{ham0}) where $S^{z}_l$ can only take the values $\pm 1$.
Then, the Hamiltonian reduces to the form,
\begin{align}
{\cal H} =& \sum_{l=1}^{L} \left[
({S}^z_{l}+{S}^z_{l+1})({\tau}^{(1)z}_{l}+{\tau}^{(2)z}_{l})+ \lambda\vector{\tau}^{(1)}_{l}\vector{\tau}^{(2)}_{l}\right]\nonumber\\
&-H\sum_{l=1}^{L}\left(\tau^{(1)z}_l+\tau^{(2)z}_l+S^z_{l}\right).
\label{hama}
\end{align}
This Hamiltonian is equivalent to Eq. (1) of Ref. \citen{str1} with
\begin{align}
\Delta\rightarrow 1,\ J_{\rm I}\rightarrow  2,\ J_{\rm H}\rightarrow\lambda,\ H_{\rm I}\rightarrow 2H,\ H_{\rm H}\rightarrow H \label{eq:replace}.
\end{align}
Following Ref.\citen{str1},  the Heisenberg spins $\vector{\tau}^{(1)}_{l}$ and $\vector{\tau}^{(2)}_{l}$ can be traced out\cite{str0} and the partition function reduces to that of the one-dimensional Ising model with $S_l^z=\pm 1$.
The free energy ${\cal G}(T,H)$ {at temperature $T$} can be obtained by the replacements (\ref{eq:replace}) 
 in the corresponding expressions for the spin-1/2 Ising-Heisenberg diamond chain in Ref. \citen{str1}.
 Entropy, specific heat, magnetization, and magnetic susceptibility are obtained by appropriate differentiations.

The possible ground states  are classified as follows:
\begin{enumerate}
\item N\'eel phase (N) : $\forall l\ T_l=1, T_l^z=\pm 1, S_l^z=-T_l^z$. This phase corresponds to the ferrimagnetic phase of Hamiltonian (1) of Ref. \citen{str1} with $S=1/2$. However, in the present model, the total magnetization vanishes. This phase belongs to a DC$\infty$ phase.
\item Paramagnetic phase (P) : $\forall l\ T_l=0$. {Ising spins} $S_l^z$ are decoupled {from each other} and can take the values $\pm 1$ independently in the absence of the magnetic field $H$. For $H > 0$, $\forall l \  S_l^z=1$. This phase corresponds to the DC0 phase.
\item Ferromagnetic phase (F) : $\forall l\ T_l^z=S_l^z=\pm 1$.
\end{enumerate}
The corresponding ground-state energies {per unit cell} are given by
$E_{\rm N}=\frac{\lambda}{4}-2,\ \
E_{\rm P}=-\frac{3}{4}\lambda-H\ \ $ and $
E_{\rm F}=\frac{\lambda}{4}+2-2H.
$
\begin{figure}
    \centering
    \includegraphics[height=3cm]{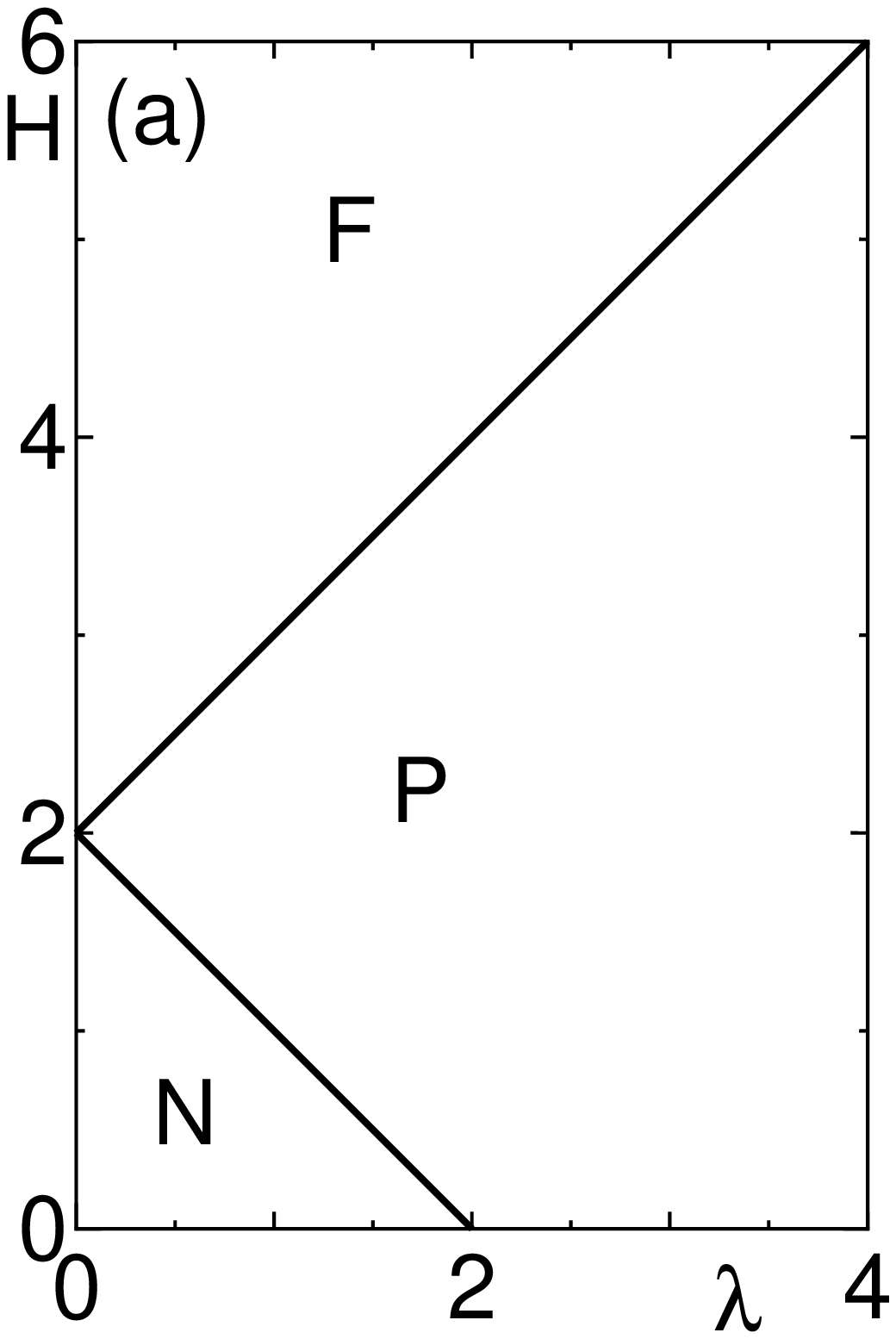} \includegraphics[height=3cm]{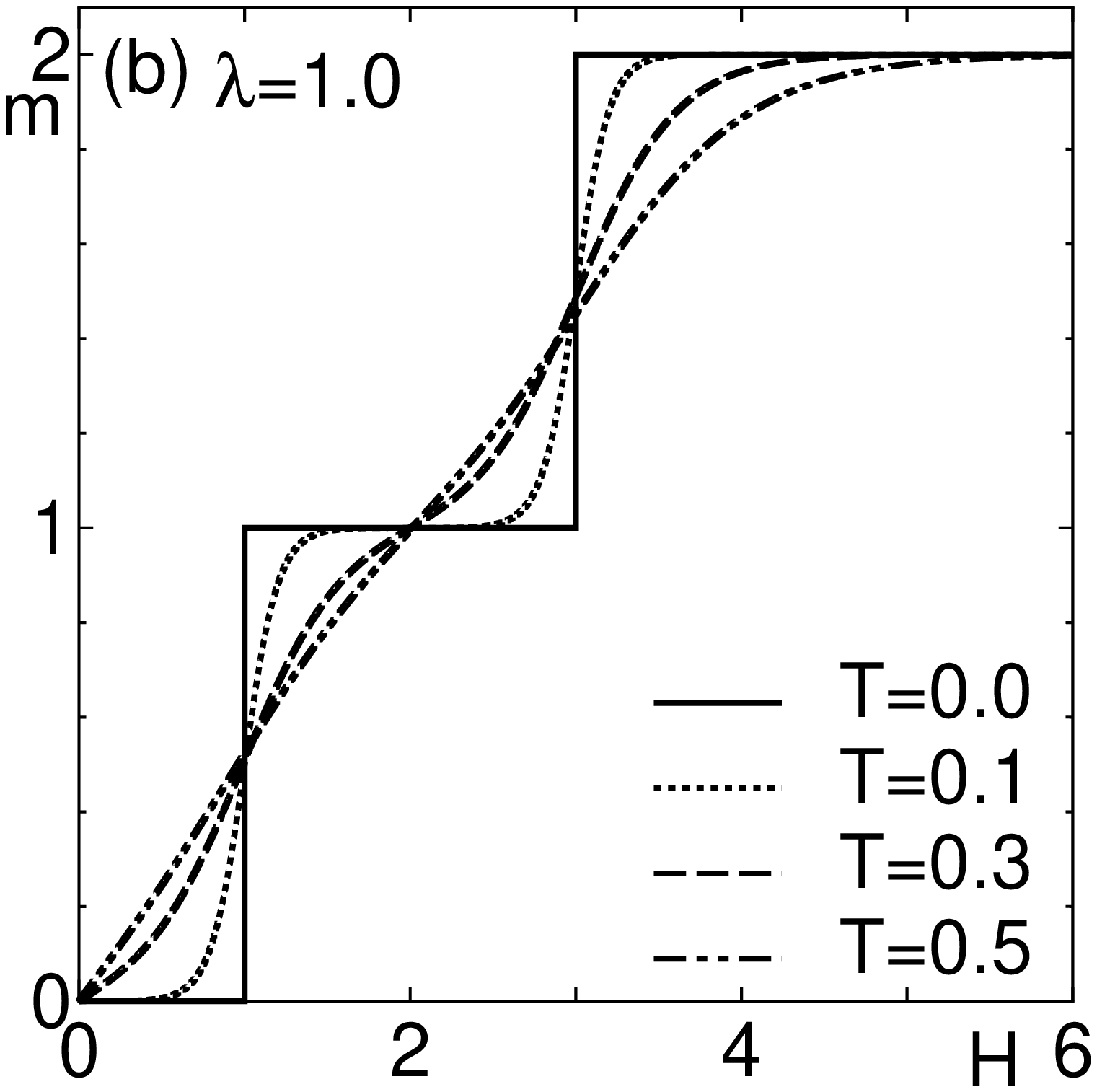} \includegraphics[height=3cm]{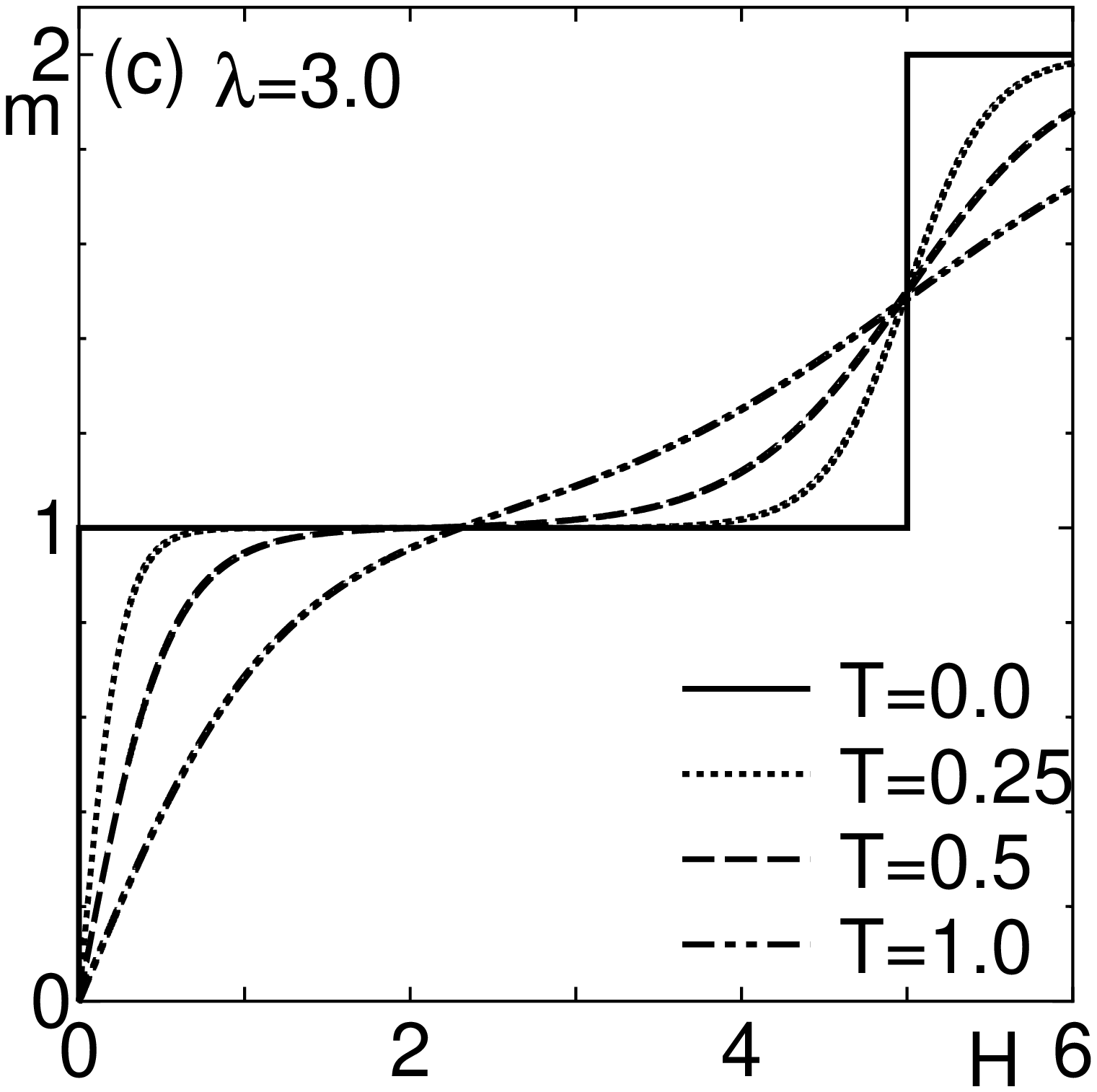}
    \caption{(a) Ground state phase diagram on the $\lambda$--$H$ plane and magnetization curves for (b) $\lambda=1$ and (c) $\lambda=3$.}
    \label{phases and mags}
\end{figure}
The phase boundaries between these phases are shown in Fig.~\ref{phases and mags}(a).
{The change of ground states with the magnetic field is clearly represented by the magnetization curves in Fig.~\ref{phases and mags}(b) and (c). The stepwise magnetization curves at $T=0$ are smeared at finite temperatures as shown in these figures.}  
It should be remarked that the P phase is induced not only by frustration ($\lambda $) but also by the magnetic field, even though all pairs $\vector{T}_l$ are nonmagnetic in the P phase and  magnetic in the N phase realized for lower magnetic fields.
\begin{figure}
    \centering
    \includegraphics[height=2.5cm]{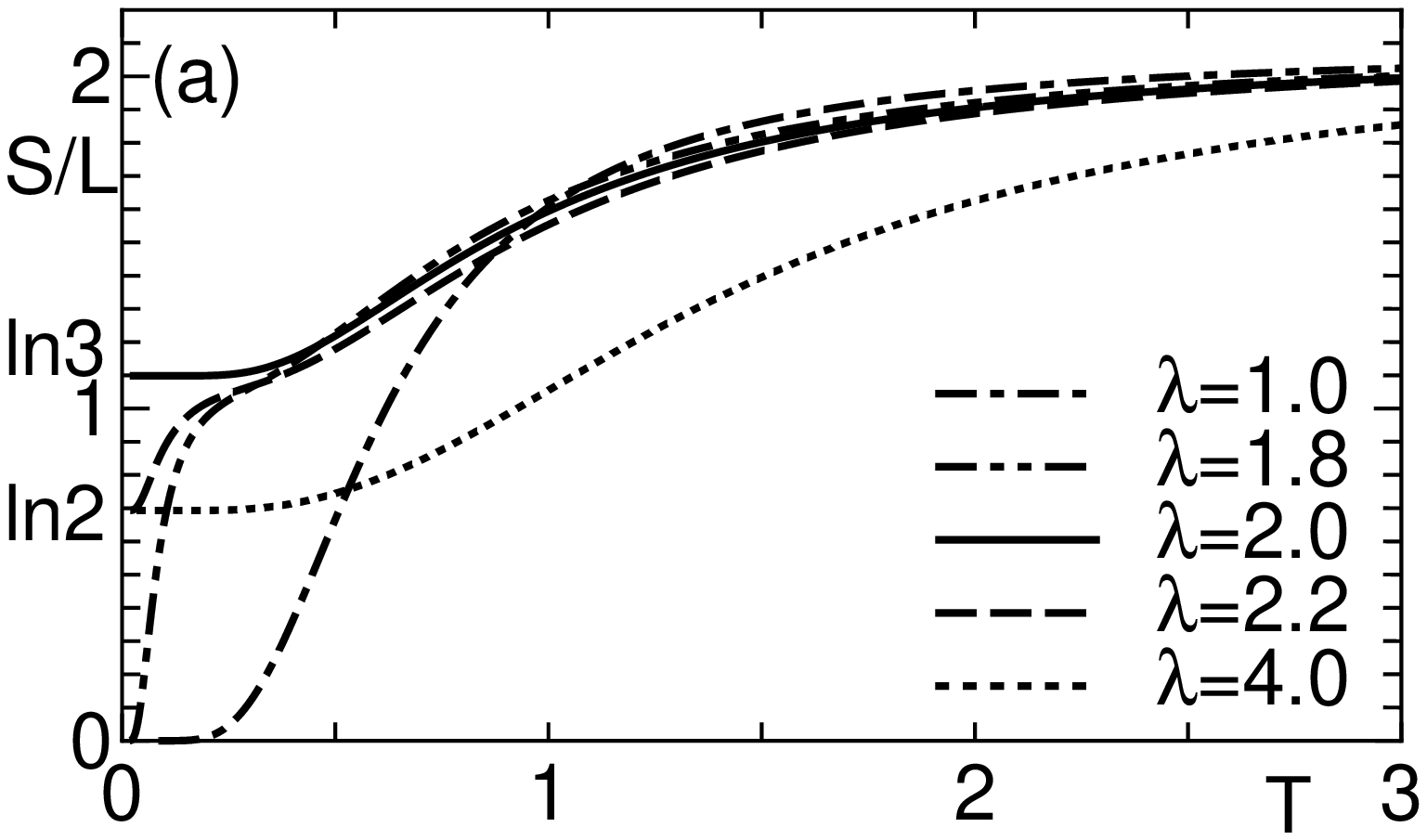} \includegraphics[height=2.5cm]{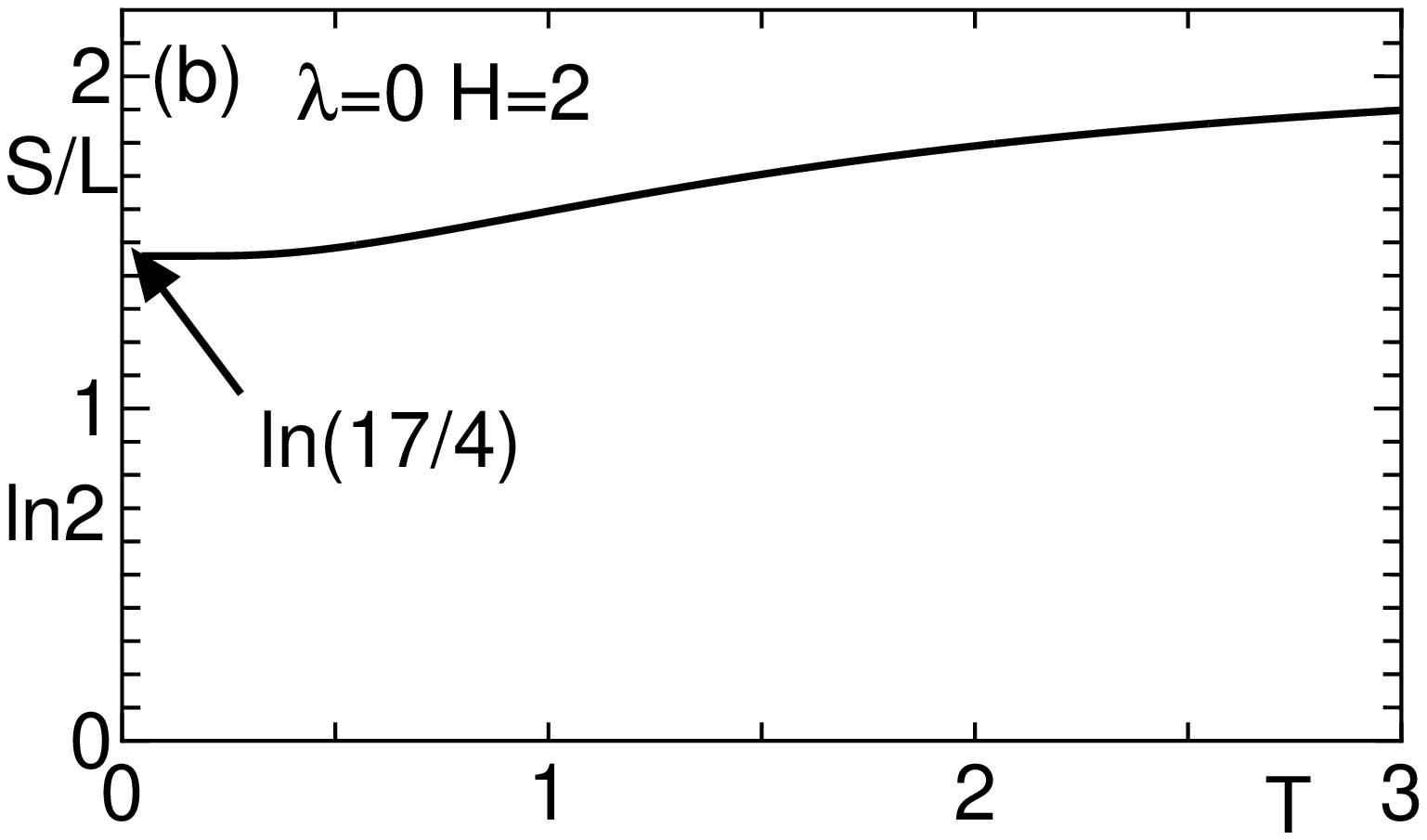}
    \caption{Temperature dependence of entropy $\entropy$ (a) for $H=0$, $1 \leq \lambda \leq 4$ and (b) $\lambda=0$, $H=2$.}
    \label{entropy_ising_1}
\vspace{-0.5cm}
\end{figure}

The temperature dependence of the entropy $\entropy$  in the absence of magnetic field is shown in Fig.~\ref{entropy_ising_1}(a) for various values of $\lambda$.
For $\lambda < 2$, the ground state is a N\'eel ordered state.
Hence, the entropy $\entropy$  vanishes as $T\rightarrow 0$.
On the other hand, for $\lambda >2$, the ground state is {a paramagnetic state} in which each $S_l^z$ can take two values.
Hence, the entropy $\entropy$ tends to $L\ln 2$  as $T\rightarrow 0$.
  At  $\lambda=2$, the creation energy of a dimer in the N\'eel background vanishes.
  Hence, an arbitrary number of dimers are present in the ground state.
  The number of configurations of $\Nd$ dimers is $_LC_{\Nd}$.
  Each segment between two dimers can take two states with total magnetization $\pm 1$.
  Hence, the number of states with $\Nd$ dimers is $_LC_{\Nd}\times 2^{\Nd}$.
  The total number of states $W$ is then given by
\begin{align}
W=\sum_{\Nd=0}^L\ _LC_{\Nd} 2^{\Nd}=(2+1)^L=3^L.
\end{align}
This gives the residual entropy $\entropy=L\ln 3$ larger than the values for $\lambda \neq 2$.
This is consistent with the data shown in Fig.~\ref{entropy_ising_1}(a).

 The temperature dependence of the entropy $\entropy$ at the triple point $(\lambda=0,H=2.0)$ is shown in Fig.~\ref{entropy_ising_1}(b).
 To estimate the residual entropy, let us consider the ground state of a diamond $\ket{S_l^z\tau^{(1)z}_l\tau^{(2)z}_lS_{l+1}^z}$.
 If $S_l^z=S_{l+1}^z=1$, the energy is equal to $-2$ irrespective of the values of  $\tau^{(1)z}_l(=\pm 1/2)$ and $\tau^{(2)z}_l(=\pm 1/2)$.
 On the other hand, for  $S_l^z=-S_{l+1}^z$ only the state with $\tau^{(1)z}_l=\tau^{(2)z}_l=1/2$ has the lowest energy $-2$. Namely, the spins $\tau^{(1)z}_l$, $\tau^{(2)z}_l$, $\tau^{(1)z}_{l-1}$, and $\tau^{(2)z}_{l-1}$ 
 on both sides of $\Ndw$ sites with $S_l^z=-1$ are fixed, while those on other $N-2\Ndw$ sites can take four possible states freely.
 The number of configurations of 
 the sites with $S_l^z=-1$ is $_LC_{\Ndw}$.
 Hence, the number of states with fixed $\Ndw$ is $_LC_{\Ndw}\times 4^{L-2\Ndw}$.
 The total number of states $W$ is then given by
\begin{align}
W=\sum_{\Ndw=0}^L\ _LC_{\Ndw} 4^{L-2\Ndw}
=\left(\frac{17}{4}\right)^L.
\end{align}
This gives the residual entropy $\entropy=L\ln (17/4)$ that is consistent with Fig.~\ref{entropy_ising_1}(b).

\begin{figure}
    \centering
    \includegraphics[height=2.5cm]{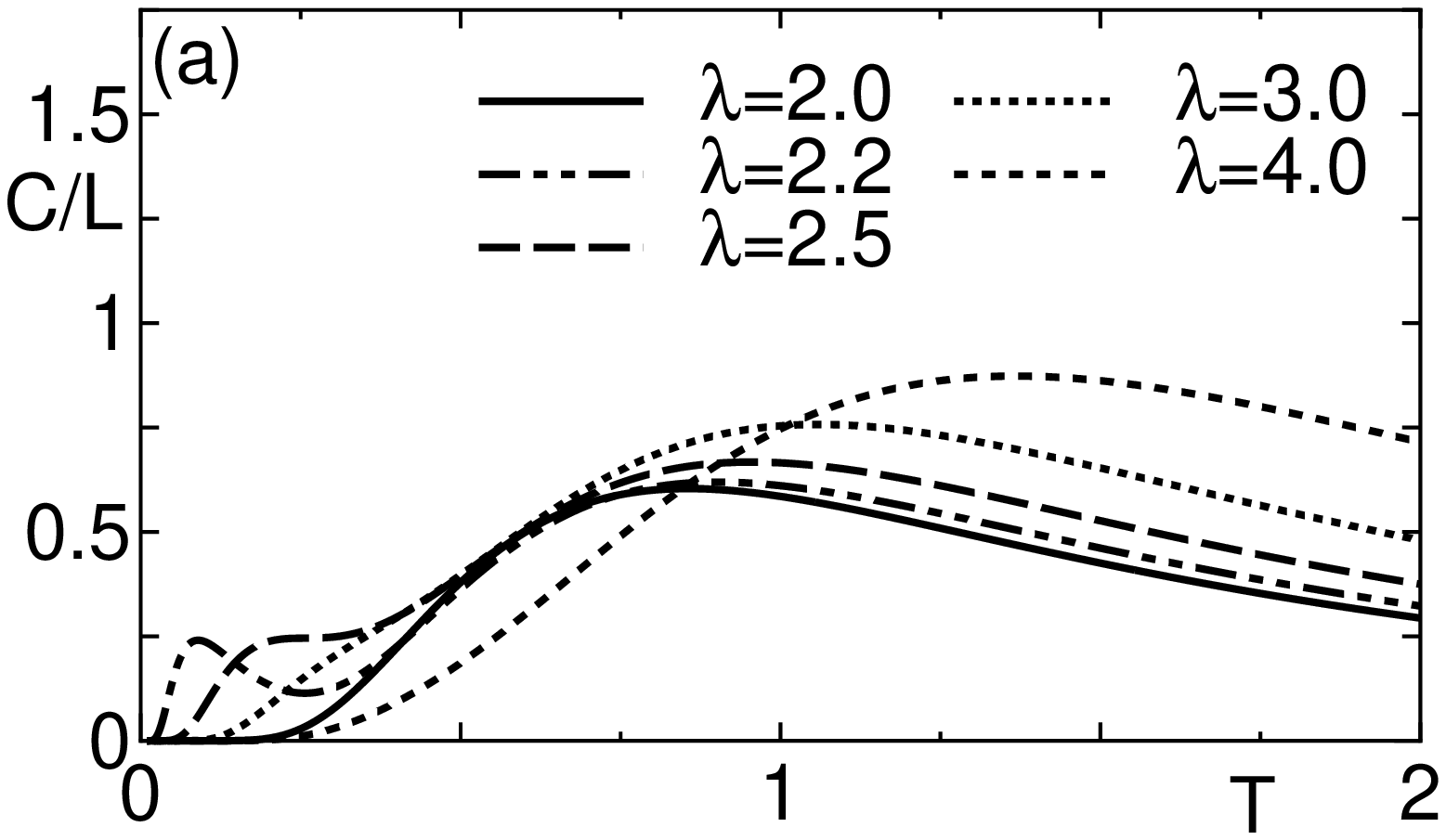}\includegraphics[height=2.5cm]{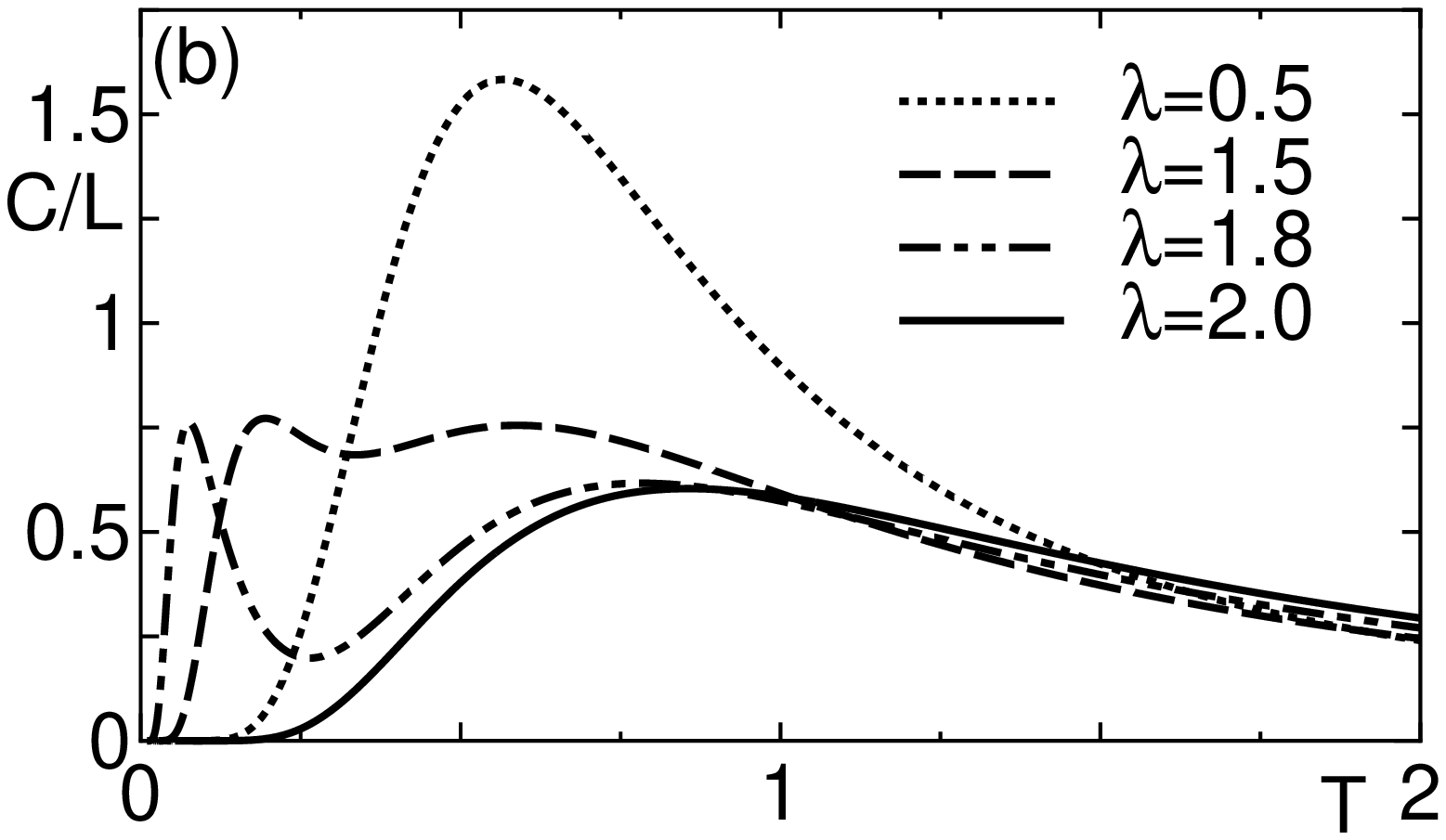}
    \caption{
        Specific heat  in the absence of magnetic field for (a) $\lambda \geq 2.0$ and (b) $\lambda \leq 2.0$.
    }\label{heat_ising_1_JH}
\vspace{-0.5cm}
  \end{figure}

The temperature dependence of the specific heat $\heat$ in the absence of magnetic field is shown in Fig.~\ref{heat_ising_1_JH}.
The low temperature peak moves to lower temperature  as $\lambda$ approaches 2.0 from both sides.
This peak disappears at $\lambda=2$.
We can interpret that the entropy released under the peak turns into the increase in the residual entropy at $\lambda=2$.
The similar phenomena are observed  
for $D=0$\cite{hida0}  at the phase boundary between DC$n$ phases with different $n$. The corresponding behavior of  entropy is observed also for finite $D$.\cite{hida1}
However, exact thermodynamics is not available in the  case of N\'eel ground state for finite $D$.
In the present strongly anisotropic limit, it is shown explicitly that this phenomenon occurs also at the DC$0$-N\'eel phase boundary.
\begin{figure}
    \centering
    \includegraphics[height=2.5cm]{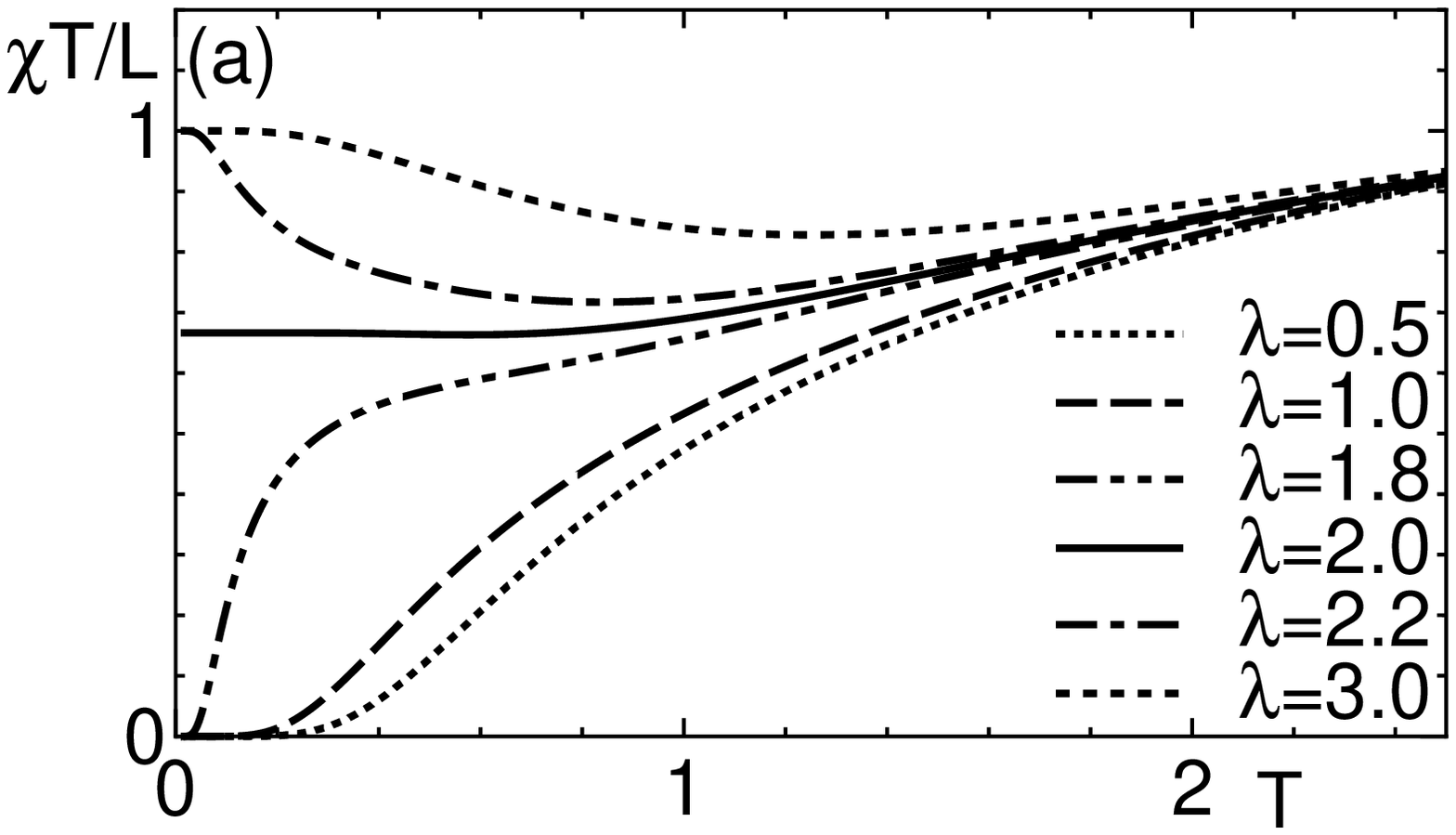}\includegraphics[height=2.5cm]{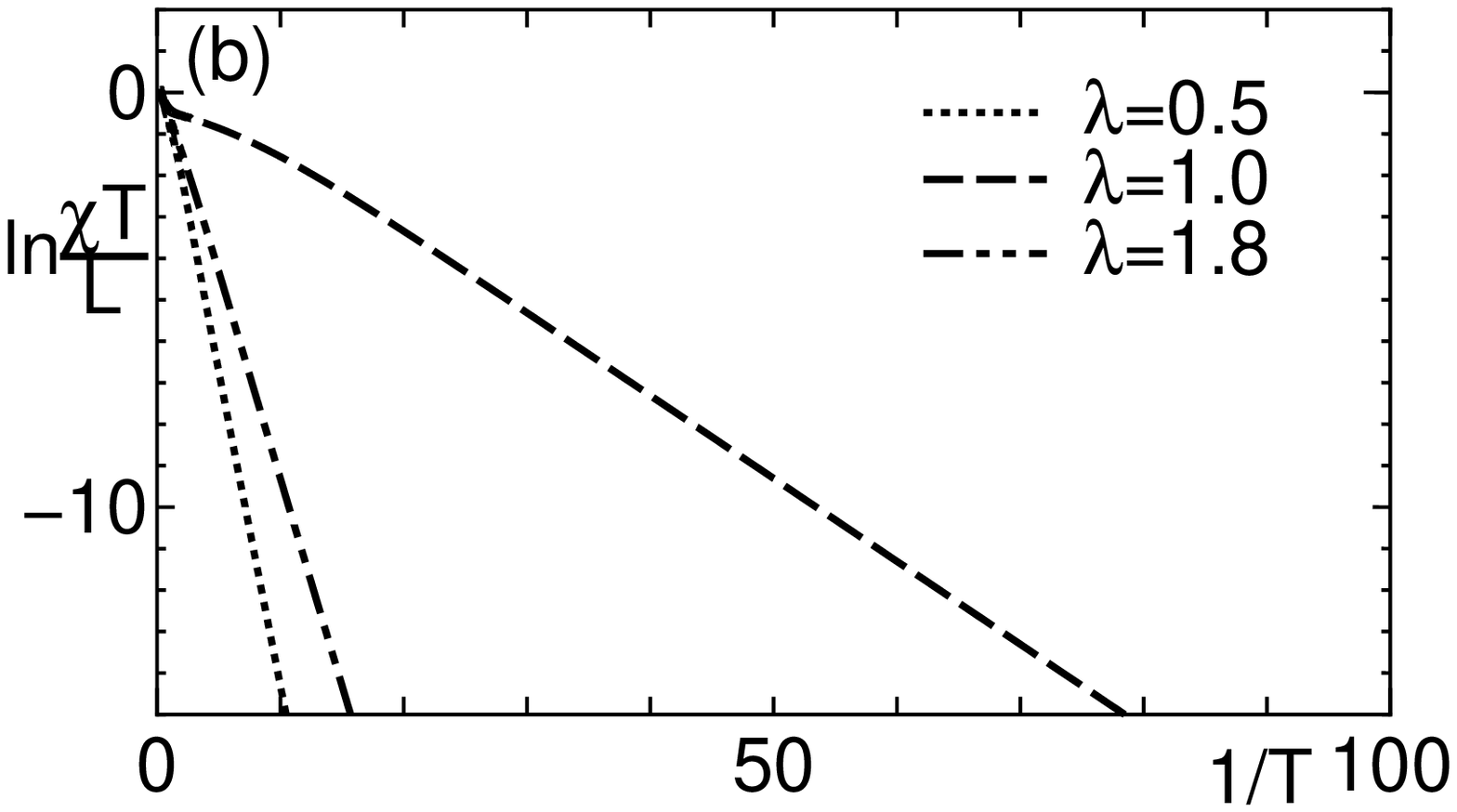}
    \caption{Temperature dependence of magnetic susceptibility $\chi$. (a) Linear plot of $\chi T/L$ against $T$.                (b) Plot of $\ln(\chi T/L)$ against $1/T$.
        }
    \label{chit_ising_1}
\vspace{-0.5cm}
  \end{figure}

The temperature dependence of the magnetic susceptibility is shown in Fig.~\ref{chit_ising_1}(a).
For $\lambda >2$, the ground state is paramagnetic and the Curie law behavior $\chi T \rightarrow L$ is observed.
This is the contribution of the spins ${S}^z_l(=\pm 1)$.
On the other hand, the susceptibility shows an exponential behavior for $\lambda < 2$ as shown in  Fig.~\ref{chit_ising_1}(b).
 However, this excitation energy is not the energy required to flip the spin $S^z_l$ or $T^z_l$ as in conventional antiferromagnets.
 The latter should be equal to $2$ irrespective of the value of $\lambda$, while the excitation energies estimated from the slopes 
of Fig.~\ref{chit_ising_1}(b) clearly depend on $\lambda$.
This can be understood in the following way: 
The excitation energy $E_{\rm d}$ of a dimer in the N\'eel background is given by
    $E_{\rm d}=\left(2-\lambda\right)$.
This implies that the number of dimers $\Nd$ 
 is proportional to $\exp \left(-\frac{\left(2-\lambda\right)}{T}\right)$.
The segment between two dimers carries an Ising spin $\pm 1$ irrespective of its length. Each segment contributes to $\chi$ by $1/T$. Under the periodic boundary condition, the number of these segments is  also equal to $\Nd$. 
Hence, we find
  \begin{align}\label{chi_syusoku}
      \chi T \propto \exp \left({-}\frac{\left(2-\lambda\right)}{T}\right).
  \end{align}
This is consistent with  the slopes of Fig.~\ref{chit_ising_1}(b).
Thus, the increase of the magnetic susceptibility with temperature in the case of N\'eel ground state results from the thermal excitation of nonmagnetic dimers.

At $\lambda=2$, the expectation value of $\Nd$ in the ground state is given by
\begin{align}
\aver{\Nd}&=\frac{1}{W}\sum_{\Nd=0}^L\Nd\ _LC_{\Nd} 2^{\Nd}=\frac{2L}{3}.
\end{align}
Hence, $\chi T/L$ tends to ${2}/{3}$ as $T\rightarrow 0$ as shown in Fig.~\ref{chit_ising_1}(a)
.

We thank S. Hoshino and H. Shinaoka for fruitful discussion and comments. This work is  supported by  JSPS KAKENHI Grant Number JP25400389.

\end{document}